*Mini review:*

# Mechanistic Models of COVID-19: Insights into Disease Progression, Vaccines, and Therapeutics

Rajat Desikan[1,‡,*], Pranesh Padmanabhan[2,‡], Andrzej M. Kierzek[1], Piet H. van der Graaf[1,3,*]

[1]Quantitative Systems Pharmacology (QSP) group, Certara, Sheffield and Canterbury, United Kingdom
[2]Clem Jones Centre for Ageing Dementia Research, Queensland Brain Institute, The University of Queensland, Brisbane, Australia
[3]Systems Biomedicine and Pharmacology, Leiden Academic Centre for Drug Research, Leiden University, Leiden, The Netherlands

[‡]Co-first authors

*Correspondence:
Dr. Rajat Desikan (rajatdesikan@gmail.com)
Prof. Piet H. van der Graaf (piet.vandergraaf@certara.com)

**Abstract**

The COVID-19 pandemic has severely impacted health systems and economies worldwide. Significant global efforts are therefore ongoing to improve vaccine efficacies, optimize vaccine deployment, and develop new antiviral therapies to combat the pandemic. Mechanistic viral dynamics and quantitative systems pharmacology models of SARS-CoV-2 infection, vaccines, immunomodulatory agents, and antiviral therapeutics have played a key role in advancing our understanding of SARS-CoV-2 pathogenesis and transmission, the interplay between innate and adaptive immunity to influence the outcomes of infection, effectiveness of treatments, mechanisms and performance of COVID-19 vaccines, and the impact of emerging SARS-CoV-2 variants. Here, we review some of the critical insights provided by these models and discuss the challenges ahead.

**1. Introduction**

COVID-19, caused by the severe acute respiratory syndrome coronavirus 2 (SARS-CoV-2) virus, has infected more than 270 million individuals worldwide, resulting in the death of over 5.3 million people as of 14th December 2021. SARS-CoV-2 is a small, single-stranded RNA virus that primarily infects the respiratory system. The clinical outcome of infection is heterogeneous, ranging from cure without symptoms to severe disease culminating in death [1]. Nearly two dozen COVID-19 vaccines have been approved for clinical use, with several hundreds of candidates in clinical and preclinical development [2]. The approved vaccines are powerful in conferring protection, particularly against severe disease [3]. Simultaneously, new antivirals such as paxlovid have been shown to significantly reduce COVID-19 deaths in clinical trials [4]. As SARS-CoV-2 continues to evolve, enormous efforts are therefore underway to understand how SARS-CoV-2 variants may impact vaccine and antiviral treatment effectiveness [5,6].

Over the past two years, SARS-CoV-2 viral dynamic models that quantitatively analyze viral load evolution in patients (Figure 1), and quantitative systems pharmacology (QSP) models that integrate systems biology and human physiology with clinical pharmacology (Figure 2*a, b*) and can thus predict the system's response to therapeutics, have been developed. These models are playing a key role in understanding SARS-CoV-2 infection as well as identifying optimal vaccine dosing regimen and antiviral treatment strategies. In this review, we first discuss insights gained into COVID-19 disease progression and the influence of host immune responses on suppressing infection. Next, we focus on modelling COVID-19 therapeutics, followed by descriptions of vaccine QSP frameworks for dose optimization and predicting efficacy. Finally, we conclude by outlining modelling efforts needed to combat the current and future pandemics.

## 2. Virus dynamics models of COVID-19 disease progression and immune response

Like most acute viral infections, COVID-19 is characterized by two phases (Figure 1*a*): a proliferation phase, where viral loads at sites of infection increase exponentially due to abundant target cells until it attains a peak, and a clearance phase, where a lack of target cells and/or mounted antiviral host immune responses lower and clear the virus from the system [7]. The relative timescales of the two phases, $\tau_p$ and $\tau_c$ (Figure 1*a*), are determined by viral infectiousness and strength of immune responses, and are signatures of transmissivity and severity of infection, therefore influencing the choice of treatment. For example, an individual in the proliferation phase may need longer isolation and closer medical monitoring due to a higher risk of complications than a person in the clearance phase. Therefore, mathematical modelling of longitudinal virus load data has been employed to not only quantitatively determine *in vivo* disease dynamics at the individual and population levels, but also the interplay between virus-host immune responses and the resulting pathology.

Statistical models are a class of models employed for quantifying SARS-CoV-2 virus dynamics, where virus load time series measured using RT-qPCR are fit to piecewise linear regression or other models to estimate individual and population virus trajectories (Figure 1*a*) [7]. While such phenomenological models were useful for deployment in the initial stages of the pandemic with limited available data, a glaring limitation is a lack of mechanistic insight into disease progression, co-evolving immune dynamics, heterogeneity of responses, and estimates of efficacies of various therapeutics and vaccines. More mechanistic virus dynamics models [8,9] (Figure 1*b*) largely address the above limitations and have provided valuable insights into the progression of many acute and chronic infections caused by HIV [8,10], hepatitis C virus (HCV)

[11], influenza [12,13], and Zika virus [14] among others. Example model fits to SARS-CoV-2 virus trajectories from two patients are shown in Figure 1*c*.

The basic model of acute viral infection, also known as the 'target cell limited model', predicts COVID-19 disease progression to be a result of interactions between SARS-CoV-2 virions, infected cells, and the availability of uninfected target cells. The effects of innate and adaptive immune responses are lumped into the viral dynamic parameters such as infection and infected cell death rate constants. In these models, after the proliferation phase, viral load declines until complete clearance due to the lack of target cells for new infection. Such models have quantitatively described and compared the within-host dynamics of MERS, SARS-CoV-1, and SARS-CoV-2, and predicted that SARS-CoV-2 had a shorter time from the symptom onset to the acute infection viral load peak compared to the other two coronaviruses [15]. Subsequent models [16] have accounted for the known delays between cellular infection and viral production by including an eclipse phase for infected cells, allowing for more accurate estimates of SARS-CoV-2 viral dynamic parameters and the within-host reproductive ratio that determines the ability of the virus to establish infection in an individual. Modelling efforts to quantify viral dynamic parameters of SARS-CoV-2 variants [17,18] of concerns and to link viral kinetics to an individual's disease transmission probability are also underway [19–21].

Unless individuals have pre-existing cross-reactive T-cell immunity to SARS-CoV-2 due to prior exposure to other infections, it takes about a week or two post-infection to mount effective T cell responses against SARS-CoV-2. Many models have added an explicit T cell compartment to understand how T cell dynamics affects SARS-CoV-2 infection and disease progression. Here, infected cells activate $CD8^+$ T cells, which in turn increases the killing rate of infected cells. This modification has been helpful to describe more rapid viral load declines observed in some individuals during the clearance phase. Recently, a minimal model consisting of the essential interactions between infected cells, $CD8^+$ T cells, and innate immune response was developed to understand the diverse outcomes of SARS-CoV-2 infection, from clearance without symptoms to severe illness followed by death [22]. By analyzing data from patients with different degrees of infection severities, the model predicted that variations in the timing and strengths of the innate and the CD8 T-cell responses could give rise to the observed spectrum of SARS-CoV-2 infection outcomes. The role of the humoral response, the other component of adaptive immunity, in resolving SARS-CoV-2 infection in non-vaccinated individuals is still unclear. By modelling and analyzing longitudinal data of viral loads and SARS-CoV-2-specific IgG and IgM antibodies in patients, a recent model suggested that the surge in IgG levels occurs 5-10 days post symptom onset and had a minor impact on viral clearance [23].

Several classes of immunomodulatory agents, such as interferons, steroids such as dexamethasone, kinase inhibitors such as baricitinib, and interleukin inhibitors such as tocilizumab, are either approved or in clinical development. The choice, timing, combination, and duration of these treatments are likely to depend on the stage and severity of the infection. For instance, in hamster models challenged with SARS-CoV-2, early prophylactic interferon treatment conferred protection whereas later treatment did not [24]. Future modelling efforts, especially larger QSP models [25–27] incorporating viral dynamics and associated innate and adaptive immune responses at various anatomical sites while accounting for host factors such as comorbidities and age, may accurately predict how different arms of the immune system would interact to determine diverse infection outcomes and guide rational optimization of treatments with immune modulators.

## 3. Models of COVID-19 therapeutics

Mathematical models combining virus dynamics with pharmacokinetic/pharmacodynamic (PK/PD) models of therapeutics have been extensively used to estimate the efficacy and identify the mechanism(s) of action of therapeutics (Figure 1*b*). Models achieve this by introducing an efficacy factor varying between 0 and 1 that lowers viral production or *de novo* infection rate constants depending on the mode of drug action [9]. For instance, Pfizer's paxlovid, which works by inhibiting the activity of SARS-CoV-2 protease [4], would lower the viral production rates in the model (Figure 1*b*). Similarly, Merck's molnupiravir, which may inhibit viral replication by increasing viral mutation rates [4], would render a fraction of progeny virions non-infectious [8,28]. Anti-SARS-CoV-2 neutralizing monoclonal antibody drugs such as REGEN-COV (casirivimab and imdevimab), sotrovimab, bamlanivimab, and etesevimab, block infection of target cells and simultaneously induce phagocytosis and clearance of virus. We expect that analyzing viral kinetics in patients treated with these antivirals in the future will: (i) quantify the treatment efficacies, (ii) clarify the *in vivo* dominant mechanism(s) of action of these antivirals, (iii) optimize dosage, as shown recently for the combination of bamlanivimab and etesevimab using population PK/PD and viral dynamics modeling [29,30], and, (iv) predict optimal antiviral combinations with high genetic barriers [8] that may lower the probability of emerging drug-resistant variants, akin to HIV and HCV combination therapies.

Several models have predicted the need for early administration before the viral peak of therapeutics that blocked new infections or lowered viral production, as most cells are likely to be infected by the time viral load peaks, and late administration is likely to inhibit only a small proportion of cellular infection [15,16,27,31]. The current estimates of the critical therapeutic efficacy, which is the minimum drug efficacy above which infection cannot be established, is ~80-95%. Until effective SARS-CoV-2 therapeutics are developed, one strategy would be to repurpose available drugs for treatment. However, analysis of SARS-CoV-2 dynamics in patients treated with different repurposed drugs revealed the efficacies were far below the critical efficacy [16]. Therefore, combination treatments that display synergistic effects are desirable [32]. We recently developed a model of SARS-CoV-2 entry into target cells and predicted that targeting two host proteases, which mediated SARS-CoV-2 entry into target cells via independent pathways, could be synergistic [33]. Such frameworks can be advantageously combined with mathematical models of drug resistance to elucidate optimal combination therapies that would simultaneously maximize synergy and minimize the probability of escape variants.

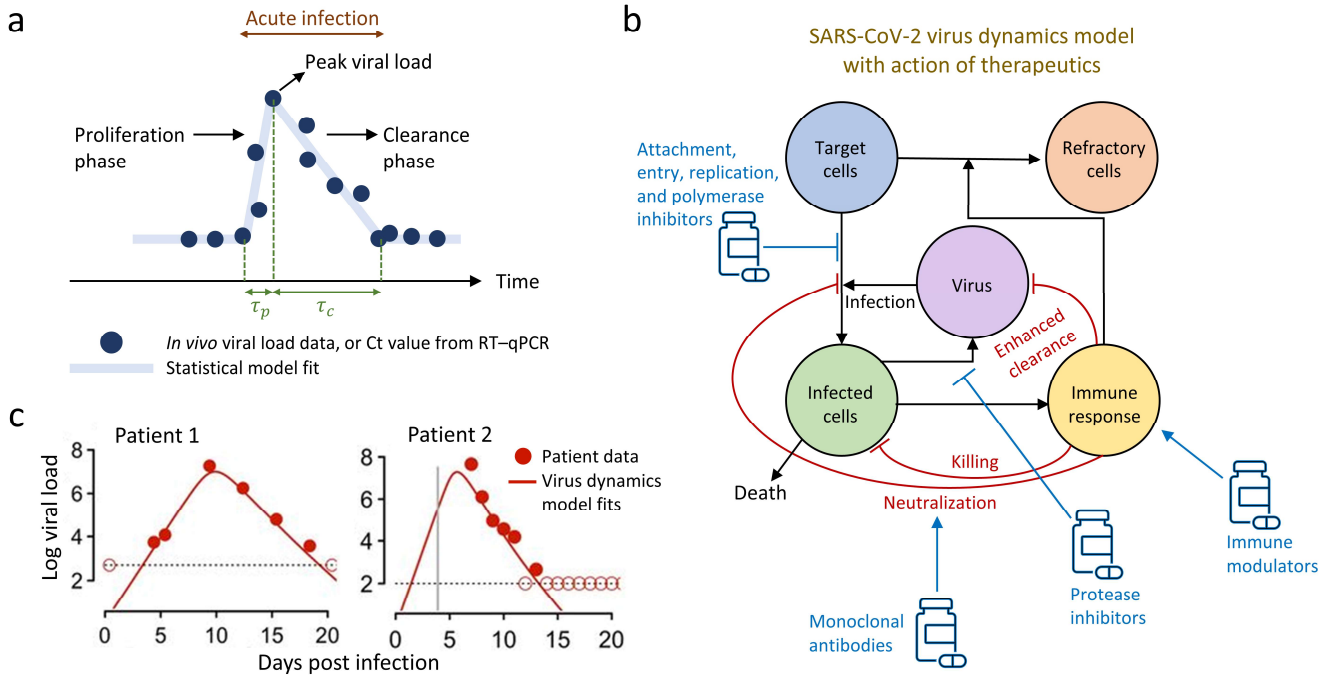

**Figure 1. Statistical and mechanistic models of SARS-CoV-2 dynamics. (a)** Schematic of predictions of a statistical model (line) together with patient data (circles) depicting the proliferation phase, $\tau_p$, and the clearance phase, $\tau_c$, of SARS-CoV-2 infection. **(b)** Schematic of a mechanistic model of within-host SARS-CoV-2 viral dynamics, describing the interactions between target cells, infected cells, virions, immune response, and immune-mediated cells refractory to infection, and the mode of action of different drugs in clinical development and application. **(c)** Representative fits with a mechanistic viral dynamics model to longitudinal viral load data from two patients; adapted from Ref. [21].

## 4. Application of QSP models for vaccine development

With a raging pandemic, no definitive sterilizing cure (yet), and substantial vulnerable populations worldwide, regulatory-approved COVID-19 vaccines are critical for safely suppressing infections and mortality. In the current scenario with many variables such as limited vaccine supplies, availability of multiple vaccine types with varying levels of protection across populations, evolving SARS-CoV-2 variants, and uncertainty regarding the durability of elicited immune responses [34], vaccine biosimulations, including with QSP [35,36], mathematical [37,38], and epidemiological [39] models, can deliver potent intelligence to support decision-making for maximal vaccine deployment. For example, early clinical trials with the Pfizer/BioNTech (NCT04368728), Moderna (NCT04470427), and Oxford-AstraZeneca (NCT04400838) vaccines were conducted with either 3-week, 4-week, or 4-6-week intervals, respectively, between the first (prime) and second (boost) doses. However, in December 2020, the UK approved a 12-week prime-boost interval to maximize the first-dose vaccine coverage across the population [40], against recommended intervals based on clinical trials and thus sparking debate among experts. Around that time, our group at Certara deployed a novel, proprietary, vaccine QSP model [35] (Figure 2*b*), which was extensively calibrated with pre-clinical and clinical mRNA vaccine data, to predict that a longer 7-8-week prime-boost interval would elicit maximal antibody titers, and reassuringly, a 12-week interval would still yield higher titers than a 3- or 4-week interval (Figure 2*c*). Strikingly, ~6 months after this prediction, the PITCH study [41] performed by employing an extended dosing interval of 6-14 weeks with the Pfizer/BioNTech vaccine quantitatively confirmed this prediction [42], highlighting the power of QSP frameworks to robustly forecast what-if scenarios in the face of uncertainty. The neutralization potential of the elicited antibodies may also be improved with delayed boost dosing (and/or lowered prime dose) due to altered B cell selection stringency in germinal centers [38]. Next, we discuss other important questions about which QSP predictions can help ameliorate uncertainty and thus significantly contribute to decision-making.

Do the elderly mount less effective immune responses upon vaccination? Extensive simulations with QSP models calibrated on age-stratified clinical trial data suggested reduced antibody titers in the elderly (65–85 years) [35]. The reduction was exacerbated at lower vaccine doses, and with time due to waning immune responses [35]. Thus, multiple boosters with current vaccine doses may be necessary for the elderly, especially with the rise of variants. On the other hand, increasing prime and boost antigen dose amounts in the elderly may elicit comparable antibody titers to that within younger individuals, suggesting that age-stratified dose optimization – higher doses for the old and lower doses for the young – may prove to be beneficial in resource-constrained settings. Further, QSP biosimulations suggest that antibody titers upon vaccination dip below protective levels within a few years across both young and old populations [35], thus perhaps necessitating periodic booster shots for all, like annual influenza vaccines.

Similar predictions with QSP models can in principle also be made for special populations where detailed clinical trials may be impractical due to recruitment challenges, or unethical, such as patients with impaired hepatic or renal systems, but for whom swift go/no-go vaccine policy decisions must be made. Such populations include neonates, pediatric (children and juveniles), pregnant and lactating women, immunocompromised adults, patients on immune-suppressive therapies, individuals with cancer and/or chronic diseases like HIV, patients who have recently undergone various surgeries, individuals on specific comedications, diabetics, and others. Since QSP biosimulations deal with computer-generated virtual patients whose characteristics and responses are constrained by mechanistic models and real-world data, they enable virtual clinical trials (VTs) on these populations. VTs are generally much faster and cost-effective compared to clinical trials, and are often used to streamline, optimize, and sometimes replace these trials. The speed factor in performing VTs and exploring myriad combinations of principal variables is pivotal for COVID-19 in the context of optimizing the vaccine supply chain and ensuring global vaccine equity. An example of such a scenario with urgent consequences for global vaccine coverage is evaluating heterologous vaccine regimen, where different vaccines are administered for the prime and subsequent booster shots. Limited clinical trials for a few vaccine combinations are available, they show promising but non-intuitive results such as an impact of the order of vaccine administration – the Oxford-AstraZeneca (prime) + Pfizer/BioNTech (booster) combination may be better than Pfizer/BioNTech + Oxford-AstraZeneca, and both mixed vaccine regimen are better than administering two doses of Oxford-AstraZeneca [43]. However, all possible vaccine combinations with varying dose and prime-boost intervals cannot be realistically tested across unvaccinated populations with large-scale clinical trials. Here, mechanistic vaccine QSP models [35] which incorporate quantitatively calibrated antigen physiological based pharmacokinetic (PBPK) modeling, thus accounting for differences across vaccine types (e.g., mRNA vaccines vs. adenovirus vector vaccines) and individual patient characteristics (Figure 2*b*), can enable rigorously performed VTs to evaluate all heterologous vaccine combinations and greatly streamline decision-making.

**5. Predicting the efficacy of COVID-19 vaccines**

What is the relationship between immune responses evoked by vaccination and protection against infection? To answer this, immune correlates of protection must be identified [44]. While cellular responses such as primed T cells undoubtedly confer some protection, neutralizing antibody (NAb) titers, usually the first line of defense against SARS-CoV-2 viruses and infection in vaccinated individuals, appears to be a robust immune correlate of vaccine efficacy from statistical analyses of clinical and epidemiological data [44–46]. Higher the NAb titers elicited by a vaccine, higher the protection accorded to the population immunized by that vaccine [45]. A mechanistic link between NAb titers and protection, which would unravel the mechanistic underpinnings of COVID-19 vaccine efficacies and enable robust efficacy predictions, was recently proposed by us [37]. First, we hypothesized that inter-individual variability in NAb responses upon vaccination can be described by assuming that the elicited NAb repertoire within an individual was a randomly sampled subset of a shape space constituting all possible NAbs. Next, we constructed such a shape space from analyses of *in vitro* dose-response neutralization curves of ~80 anti-SARS-CoV-2 NAbs and recapitulated NAb repertoires of convalescent patients by sampling from that subset, thus validating the shape space. Finally, we developed a within-host model of SARS-CoV-2 infection which quantitatively captured human viral kinetics *in vivo*, coupled individual NAb responses after vaccination (sampled from the shape space) to virus dynamics, performed VTs of vaccinated populations with a spectrum of immune and virus characteristics, and predicted vaccine efficacies. Our predictions quantitatively captured the efficacies of major vaccines from clinical trial data and offers a framework for assessing and including other immune correlates of protection, thus opening avenues for optimizing vaccine deployment.

## 6. Outlook: Combining models of within-host disease progression, therapeutics, and vaccines with population epidemiological models

We anticipate that the optimal development and deployment of therapeutics and vaccines, especially in times of urgent need such as global pandemics, would be greatly accelerated by integrated, semi-mechanistic/statistical, population-based models. Such multiscale frameworks would seamlessly combine models of basic vaccine immunobiology [38,47], individual-scale infection models including within-host models of pathogen-immune dynamics and disease progression [25,26], action of anti-pathogen and immunomodulatory therapeutics, vaccine QSP models accounting for antigen PBPK and inter-individual variations [35–37,48], and population epidemiological models [39,49,50] that would enable real-world efficacy-predictions in the context of spreading and evolving disease as a function of virus/patient characteristics, social policies, resource constraints, and human behavior. The wealth of fine-resolution and diverse data generated by massive consortium efforts during the COVID-19 pandemic offer a unique opportunity to calibrate and validate such mega-models. Subsequently, these frameworks may enable static or dynamic predictions of waning anamnestic immune responses and protection against future disease while accounting for inter-vaccine differences, history of infection and vaccination in convalescent individuals, heterologous vaccinations, special populations, individual genetics, pharmacogenomics, combination therapies in vaccinated and unvaccinated individuals, and epidemiological aspects such as optimal supply/deployment of vaccines in the face of resource constraints and rapidly transmitting disease, pathogen evolution, impact of interventional measures such as lockdowns and vaccination strategies, and changes in human behavior. Such simulations may enable both 'nowcasting' and forecasting in the face of uncertainty, and better guide not only drug/vaccine discovery and dose optimization, but also holistic policies towards the appropriate administration of vaccines and therapeutics for effectively combating future diseases and pandemics.

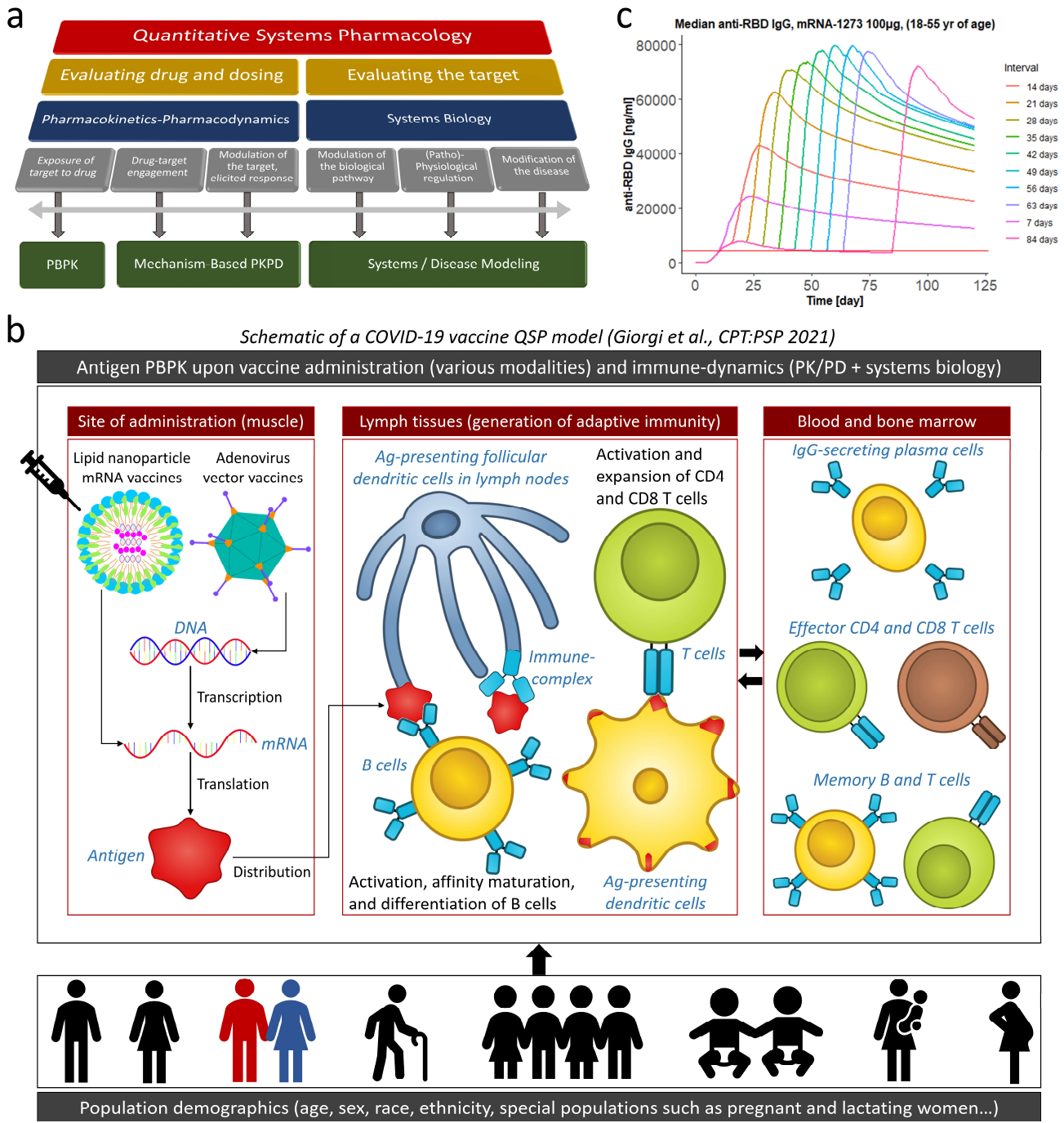

**Figure 2. Mechanistic QSP models of SARS-CoV-2 vaccines. (a)** Defining the scope of QSP modelling frameworks; adapted from Ref. [51]. **(b)** Schematic of the Certara vaccine simulator, a QSP model incorporating antigen physiologically-based pharmacokinetics upon either mRNA or adenovirus vaccine administration, immune-dynamics and resulting antibody and cellular responses, and population attributes including age, sex, race, ethnicity, HLA-genetics, and special populations such as immunocompromised, pregnant, and lactating women. (More details

in [35]). **(c)** Predictions of the optimal prime-boost interval based on IgG responses upon Moderna vaccine administration; adapted from Ref. [35].

## Acknowledgement

We thank Prof. Narendra M. Dixit for discussions and critically evaluating our manuscript.

## Author contributions

All authors were involved in conceptualizing the study, analysis, interpretation of data, and revisions. Writing of the first draft: R.D. and P.P. All authors have given final approval of the manuscript. Equal contributions: R.D. and P.P.

## Declaration of interest

R.D., A.M.K., and P.H.v.d.G. are employees of Certara QSP and were involved in the development of Certara's vaccine QSP model. No funding was received by the authors for this work. Ethical approval was not required for this work. All the authors declare no conflict of interest.

## References

[1] Brodin P. Immune determinants of COVID-19 disease presentation and severity. Nat Med 2021;27:28–33. https://doi.org/10.1038/s41591-020-01202-8.
[2] Mathieu E, Ritchie H, Ortiz-Ospina E, Roser M, Hasell J, Appel C, et al. A global database of COVID-19 vaccinations. Nat Hum Behav 2021;5:947–53. https://doi.org/10.1038/s41562-021-01122-8.
[3] McDonald I, Murray SM, Reynolds CJ, Altmann DM, Boyton RJ. Comparative systematic review and meta-analysis of reactogenicity, immunogenicity and efficacy of vaccines against SARS-CoV-2. Npj Vaccines 2021;6:1–14. https://doi.org/10.1038/s41541-021-00336-1.
[4] Ledford H. COVID antiviral pills: what scientists still want to know. Nature 2021;599:358–9. https://doi.org/10.1038/d41586-021-03074-5.

[5] Tao K, Tzou PL, Nouhin J, Gupta RK, de Oliveira T, Kosakovsky Pond SL, et al. The biological and clinical significance of emerging SARS-CoV-2 variants. Nat Rev Genet 2021;22:757–73. https://doi.org/10.1038/s41576-021-00408-x.
[6] Harvey WT, Carabelli AM, Jackson B, Gupta RK, Thomson EC, Harrison EM, et al. SARS-CoV-2 variants, spike mutations and immune escape. Nat Rev Microbiol 2021;19:409–24. https://doi.org/10.1038/s41579-021-00573-0.
[7] Kissler SM, Fauver JR, Mack C, Olesen SW, Tai C, Shiue KY, et al. Viral dynamics of acute SARS-CoV-2 infection and applications to diagnostic and public health strategies. PLOS Biology 2021;19:e3001333. https://doi.org/10.1371/journal.pbio.3001333.
[8] Perelson AS, Ke R. Mechanistic Modeling of SARS-CoV-2 and Other Infectious Diseases and the Effects of Therapeutics. Clinical Pharmacology & Therapeutics 2021;109:829–40. https://doi.org/10.1002/cpt.2160.
[9] Padmanabhan P, Dixit NM. Models of Viral Population Dynamics. In: Domingo E, Schuster P, editors. Quasispecies: From Theory to Experimental Systems, vol. 392, Cham: Springer International Publishing; 2015, p. 277–302. https://doi.org/10.1007/82_2015_458.
[10] Desikan R, Raja R, Dixit NM. Early exposure to broadly neutralizing antibodies may trigger a dynamical switch from progressive disease to lasting control of SHIV infection. PLoS Comput Biol 2020;16:e1008064. https://doi.org/10.1371/journal.pcbi.1008064.
[11] Padmanabhan P, Garaigorta U, Dixit NM. Emergent properties of the interferon-signalling network may underlie the success of hepatitis C treatment. Nat Commun 2014;5:3872. https://doi.org/10.1038/ncomms4872.
[12] Myers MA, Smith AP, Lane LC, Moquin DJ, Vogel P, Woolard S, et al. Dynamically Linking Influenza Virus Infection with Lung Injury to Predict Disease Severity. Systems Biology; 2019. https://doi.org/10.1101/555276.
[13] Baccam P, Beauchemin C, Macken CA, Hayden FG, Perelson AS. Kinetics of Influenza A Virus Infection in Humans. Journal of Virology 2006;80:7590–9. https://doi.org/10.1128/JVI.01623-05.
[14] Best K, Guedj J, Madelain V, Lamballerie X de, Lim S-Y, Osuna CE, et al. Zika plasma viral dynamics in nonhuman primates provides insights into early infection and antiviral strategies. PNAS 2017;114:8847–52. https://doi.org/10.1073/pnas.1704011114.
[15] Kim KS, Ejima K, Iwanami S, Fujita Y, Ohashi H, Koizumi Y, et al. A quantitative model used to compare within-host SARS-CoV-2, MERS-CoV, and SARS-CoV dynamics provides insights into the pathogenesis and treatment of SARS-CoV-2. PLOS Biology 2021;19:e3001128. https://doi.org/10.1371/journal.pbio.3001128.
[16] Gonçalves A, Bertrand J, Ke R, Comets E, de Lamballerie X, Malvy D, et al. Timing of Antiviral Treatment Initiation is Critical to Reduce SARS-CoV-2 Viral Load. CPT: Pharmacometrics & Systems Pharmacology 2020;9:509–14. https://doi.org/10.1002/psp4.12543.
[17] Ke R, Martinez PP, Smith RL, Gibson LL, Mirza A, Conte M, et al. Daily sampling of early SARS-CoV-2 infection reveals substantial heterogeneity in infectiousness. 2021. https://doi.org/10.1101/2021.07.12.21260208.
[18] Kissler SM, Fauver JR, Mack C, Tai CG, Breban MI, Watkins AE, et al. Viral dynamics of SARS-CoV-2 variants in vaccinated and unvaccinated individuals. 2021. https://doi.org/10.1101/2021.02.16.21251535.
[19] Marc A, Kerioui M, Blanquart F, Bertrand J, Mitjà O, Corbacho-Monné M, et al. Quantifying the relationship between SARS-CoV-2 viral load and infectiousness. ELife 2021;10:e69302. https://doi.org/10.7554/eLife.69302.
[20] Goyal A, Reeves DB, Cardozo-Ojeda EF, Schiffer JT, Mayer BT. Viral load and contact heterogeneity predict SARS-CoV-2 transmission and super-spreading events. ELife 2021;10:e63537. https://doi.org/10.7554/eLife.63537.
[21] Ke R, Zitzmann C, Ho DD, Ribeiro RM, Perelson AS. In vivo kinetics of SARS-CoV-2 infection and its relationship with a person's infectiousness. PNAS 2021;118. https://doi.org/10.1073/pnas.2111477118.

[22] Chatterjee B, Sandhu HS, Dixit NM. The relative strength and timing of innate immune and CD8 T-cell responses underlie the heterogeneous outcomes of SARS-CoV-2 infection. 2021. https://doi.org/10.1101/2021.06.15.21258935.

[23] Yang S, Jerome KR, Greninger AL, Schiffer JT, Goyal A. Endogenously Produced SARS-CoV-2 Specific IgG Antibodies May Have a Limited Impact on Clearing Nasal Shedding of Virus during Primary Infection in Humans. Viruses 2021;13:516. https://doi.org/10.3390/v13030516.

[24] Bessière P, Wasniewski M, Picard-Meyer E, Servat A, Figueroa T, Foret-Lucas C, et al. Intranasal type I interferon treatment is beneficial only when administered before clinical signs onset in the SARS-CoV-2 hamster model. PLOS Pathogens 2021;17:e1009427. https://doi.org/10.1371/journal.ppat.1009427.

[25] Dai W, Rao R, Sher A, Tania N, Musante CJ, Allen R. A Prototype QSP Model of the Immune Response to SARS-CoV-2 for Community Development. CPT Pharmacometrics Syst Pharmacol 2021;10:18–29. https://doi.org/10.1002/psp4.12574.

[26] Voutouri C, Nikmaneshi MR, Hardin CC, Patel AB, Verma A, Khandekar MJ, et al. In silico dynamics of COVID-19 phenotypes for optimizing clinical management. PNAS 2021;118. https://doi.org/10.1073/pnas.2021642118.

[27] Rao R, Musante CJ, Allen R. A Quantitative Systems Pharmacology Model of the Pathophysiology and Treatment of COVID-19 Predicts Optimal Timing of Pharmacological Interventions. 2021. https://doi.org/10.1101/2021.12.07.21267277.

[28] Dixit NM, Layden-Almer JE, Layden TJ, Perelson AS. Modelling how ribavirin improves interferon response rates in hepatitis C virus infection. Nature 2004;432:922–4. https://doi.org/10.1038/nature03153.

[29] Chigutsa E, O'Brien L, Ferguson-Sells L, Long A, Chien J. Population Pharmacokinetics and Pharmacodynamics of the Neutralizing Antibodies Bamlanivimab and Etesevimab in Patients With Mild to Moderate COVID-19 Infection. Clinical Pharmacology & Therapeutics 2021;110:1302–10. https://doi.org/10.1002/cpt.2420.

[30] Venkatakrishnan K, van der Graaf PH. Model-Informed Drug Development: Connecting the Dots With a Totality of Evidence Mindset to Advance Therapeutics. Clinical Pharmacology & Therapeutics 2021;110:1147–54. https://doi.org/10.1002/cpt.2422.

[31] Goyal A, Cardozo-Ojeda EF, Schiffer JT. Potency and timing of antiviral therapy as determinants of duration of SARS-CoV-2 shedding and intensity of inflammatory response. Science Advances n.d.;6:eabc7112. https://doi.org/10.1126/sciadv.abc7112.

[32] Padmanabhan P, Dixit N. Modeling suggests a mechanism of synergy between hepatitis c virus entry inhibitors and drugs of other classes. CPT:PSP 2015;4:445–53. https://doi.org/10.1002/psp4.12005.

[33] Padmanabhan P, Desikan R, Dixit NM. Targeting TMPRSS2 and Cathepsin B/L together may be synergistic against SARS-CoV-2 infection. PLOS Computational Biology 2020;16:e1008461. https://doi.org/10.1371/journal.pcbi.1008461.

[34] Kim JH, Marks F, Clemens JD. Looking beyond COVID-19 vaccine phase 3 trials. Nat Med 2021;27:205–11. https://doi.org/10.1038/s41591-021-01230-y.

[35] Giorgi M, Desikan R, Graaf PH, Kierzek AM. Application of quantitative systems pharmacology to guide the optimal dosing of COVID-19 vaccines. CPT Pharmacometrics Syst Pharmacol 2021:psp4.12700. https://doi.org/10.1002/psp4.12700.

[36] Selvaggio G, Leonardelli L, Lofano G, Fresnay S, Parolo S, Medini D, et al. A quantitative systems pharmacology approach to support mRNA vaccine development and optimization. CPT: Pharmacometrics & Systems Pharmacology n.d.;n/a. https://doi.org/10.1002/psp4.12721.

[37] Padmanabhan P, Desikan R, Dixit NM. Modelling the population-level protection conferred by COVID-19 vaccination. 2021. https://doi.org/10.1101/2021.03.16.21253742.

[38] Garg AK, Mittal S, Padmanabhan P, Desikan R, Dixit NM. Increased B Cell Selection Stringency In Germinal Centers Can Explain Improved COVID-19 Vaccine Efficacies With Low Dose Prime or Delayed Boost. Frontiers in Immunology 2021;12:5064. https://doi.org/10.3389/fimmu.2021.776933.

[39] Saad-Roy CM, Morris SE, Metcalf CJE, Mina MJ, Baker RE, Farrar J, et al. Epidemiological and evolutionary considerations of SARS-CoV-2 vaccine dosing regimes. Science 2021;372:363–70. https://doi.org/10.1126/science.abg8663.

[40] Iacobucci G, Mahase E. Covid-19 vaccination: What's the evidence for extending the dosing interval? BMJ 2021;372:n18. https://doi.org/10.1136/bmj.n18.

[41] Payne RP, Longet S, Austin JA, Skelly D, Dejnirattisai W, Adele S, et al. Sustained T Cell Immunity, Protection and Boosting Using Extended Dosing Intervals of BNT162b2 mRNA Vaccine. Rochester, NY: Social Science Research Network; 2021. https://doi.org/10.2139/ssrn.3891065.

[42] Certara's Vaccine Simulator™ Accurately Predicted Optimal Timing Between Doses for COVID-19 | Certara, Inc. n.d. https://ir.certara.com/news-releases/news-release-details/certaras-vaccine-simulatortm-accurately-predicted-optimal-timing/ (accessed October 3, 2021).

[43] Liu X, Shaw RH, Stuart ASV, Greenland M, Aley PK, Andrews NJ, et al. Safety and immunogenicity of heterologous versus homologous prime-boost schedules with an adenoviral vectored and mRNA COVID-19 vaccine (Com-COV): a single-blind, randomised, non-inferiority trial. The Lancet 2021;398:856–69. https://doi.org/10.1016/S0140-6736(21)01694-9.

[44] Krammer F. A correlate of protection for SARS-CoV-2 vaccines is urgently needed. Nat Med 2021;27:1147–8. https://doi.org/10.1038/s41591-021-01432-4.

[45] Khoury DS, Cromer D, Reynaldi A, Schlub TE, Wheatley AK, Juno JA, et al. Neutralizing antibody levels are highly predictive of immune protection from symptomatic SARS-CoV-2 infection. Nat Med 2021;27:1205–11. https://doi.org/10.1038/s41591-021-01377-8.

[46] Earle KA, Ambrosino DM, Fiore-Gartland A, Goldblatt D, Gilbert PB, Siber GR, et al. Evidence for antibody as a protective correlate for COVID-19 vaccines. Vaccine 2021;39:4423–8. https://doi.org/10.1016/j.vaccine.2021.05.063.

[47] Garg AK, Desikan R, Dixit NM. Preferential presentation of high-affinity immune complexes in germinal centers can explain how passive immunization improves the humoral response. Cell Reports 2019;29:3946–57.

[48] Chen X, Hickling T, Vicini P. A Mechanistic, Multiscale Mathematical Model of Immunogenicity for Therapeutic Proteins: Part 1-Theoretical Model. CPT: Pharmacometrics & Systems Pharmacology 2014;3:133. https://doi.org/10.1038/psp.2014.30.

[49] Giordano G, Colaneri M, Di Filippo A, Blanchini F, Bolzern P, De Nicolao G, et al. Modeling vaccination rollouts, SARS-CoV-2 variants and the requirement for non-pharmaceutical interventions in Italy. Nat Med 2021;27:993–8. https://doi.org/10.1038/s41591-021-01334-5.

[50] Jentsch PC, Anand M, Bauch CT. Prioritising COVID-19 vaccination in changing social and epidemiological landscapes: a mathematical modelling study. The Lancet Infectious Diseases 2021;21:1097–106. https://doi.org/10.1016/S1473-3099(21)00057-8.

[51] Vicini P, van der Graaf PH. Systems Pharmacology for Drug Discovery and Development: Paradigm Shift or Flash in the Pan? Clinical Pharmacology & Therapeutics 2013;93:379–81. https://doi.org/10.1038/clpt.2013.40.